\documentclass{emulateapj}
\usepackage{amsmath}

\newcommand{\lea}{{\>\rlap{\raise2pt\hbox{$<$}}\lower3pt\hbox{$\sim$} \>}}
\newcommand{\gea}{{\>\rlap{\raise2pt\hbox{$>$}}\lower3pt\hbox{$\sim$} \>}}
\newcommand{\hii}{H~\textsc{ii}}

\newcommand{\msun}{M_{\odot}}
\newcommand{\lsun}{L_{\odot}}

\begin{document}

\title{Stellar Feedback in Molecular Clouds and its Influence on 
the Mass Function of Young Star Clusters}

\author{S. Michael Fall\altaffilmark{1}, 
        Mark R. Krumholz\altaffilmark{2}, and 
        Christopher D. Matzner\altaffilmark{3}}


\slugcomment{Accepted to ApJ}

\altaffiltext{1}{Space Telescope Science Institute,
         3700 San Martin Drive, Baltimore, MD 21218;
         fall@stsci.edu}
\altaffiltext{2}{Department of Astronomy and Astrophysics,
         University of California, Santa Cruz, CA 95064;
         krumholz@ucolick.org}
\altaffiltext{3}{Department of Astronomy and Astrophysics,
         University of Toronto, Toronto, ON M5S 3H8, Canada;
         matzner@astro.utoronto.ca}

\begin{abstract}

We investigate how the removal of interstellar material by stellar 
feedback limits the efficiency of star formation in molecular clouds 
and how this determines the shape of the mass function of young star 
clusters.
In particular, we derive relations between the power-law exponents of
the mass functions of the clouds and clusters in the limiting regimes
in which the feedback is energy-driven and momentum-driven, corresponding 
to minimum and maximum radiative losses, and likely to bracket all
realistic cases. 
We find good agreement between the predicted and observed exponents,
especially for momentum-driven feedback, provided the protoclusters
have roughly constant mean surface density, as 
indicated by observations of the star-forming clumps within
molecular clouds.
We also consider a variety of specific feedback mechanisms, concluding
that \hii\ regions inflated by radiation pressure predominate in massive 
protoclusters, a momentum-limited process when photons can escape after
only a few interactions with dust grains.
We show in this case that the star formation efficiency depends on the 
masses and sizes of the protoclusters only through their mean surface 
density, thus ensuring consistency between the observed exponents of 
the mass functions of the clouds and clusters. 
Our numerical estimate of this efficiency is also consistent with
observations. 

\end{abstract}


\section{Introduction}

Most stars form in protoclusters in dense molecular clumps 
(\citealt{lada03}; \citealt{mckee07b}).
The energy and momentum injected by young stars then removes 
the remaining interstellar material (ISM), thus ending further 
star formation and reducing the gravitational binding energy 
of the protoclusters. 
This feedback limits the efficiency of star formation---the 
ratio of final stellar mass to initial interstellar mass---to 
only $20-30\%$, and leaves many protoclusters unbound, with 
their constituent stars free to disperse. 
Even those protoclusters that survive will lose some stars by
ISM removal and subsequent processes.

Two of the best probes of these formation and disruption
processes are the mass functions of molecular clouds and
young star clusters, defined as the number of objects 
per unit mass, $\psi(M) \equiv dN/dM$.
For molecular clouds, the best-studied galaxies are the
Milky Way and the Large Magellanic Cloud (LMC), while for
star clusters, they are the Antennae and the LMC.
In these and other cases, the observed mass functions can 
be represented by power laws, $\psi(M) \propto M^{\beta}$,
from $10^4M_{\odot}$ or below to $10^6M_{\odot}$ or above.
Giant molecular clouds (GMCs) identified in CO surveys have 
$\beta \approx -1.7$ \citep{rosolowsky05b, blitz07a, fukui08a}.
This exponent is also found for massive self-gravitating clumps 
within GMCs, the formation sites of star clusters, whether they 
are identified by CO emission \citep{bertoldi92} or higher-density 
tracers such as C$^{18}$O, $^{13}$CO, and thermal dust emission 
\citep{reid06b, munoz07a, wong08a}.
Young star clusters have $\beta \approx -2.0$  
\citep{elmegreen97a, mckee97, zhang99b, dowell08a, 
fall09a, chandar09a}. 
The similar exponents for clouds and clusters indicate 
that the efficiency of star formation and probability 
of disruption are at most weak functions of mass.
This conclusion is reinforced by the fact that $\beta$ 
is the same for $10^7-10^8$ yr-old clusters as it is for 
$10^6-10^7$ yr-old clusters \citep{zhang99b, fall09a, 
chandar09a}.
 
These empirical results may at first seem puzzling.
Low-mass protoclusters have lower binding energy per 
unit mass and should therefore be easier to disrupt 
than high-mass protoclusters.
Indeed, several authors have proposed that feedback would 
cause a bend in the mass function of young clusters at 
$M \sim 10^5M_{\odot}$, motivated in part by the well-known
turnover in the mass function of old globular clusters 
\citep{kroupa02a, baumgardt08a, parmentier08a}.
For young clusters, such a feature is not observed (as noted 
above), while for globular clusters, it arises from almost
any initial conditions as a consequence of stellar escape 
driven by two-body relaxation over $\sim 10^{10}$~yr
\citep[and references therein]{fall01a, mclaughlin08a}.
Nevertheless, we are left with an important question: 
What are the physical reasons for the observed similarity 
of the mass functions of molecular clouds and young star 
clusters? 

The goal of this Letter is to answer this question. 
In Section~\ref{energymomentum}, we derive some general  
relations between the mass functions of clouds and 
clusters. 
In Section~\ref{sec:efficiency}, we review a variety of 
specific feedback processes and estimate the star formation 
efficiency for radiation pressure, the dominant process in 
massive, compact protoclusters. 
We summarize in Section~\ref{sec:conclusion}.
 
\section{Mass Functions}
\label{energymomentum}

The radiative losses inside protoclusters determine how much of 
the energy input by stellar feedback is available for ISM removal. 
This in turn depends on the cloud structure and the specific 
feedback mechanisms involved, but two limiting regimes bracket 
all realistic situations: energy-driven, with no radiative 
losses, and momentum-driven, with maximum radiative losses.
We estimate the mass of stars $M_*$ and the corresponding
efficiency of star formation, ${\cal E} = M_*/M$, needed to 
remove the ISM from protoclusters in these regimes as follows.
We characterize a protocluster by its mass $M$, half-mass radius 
$R_h$, mean surface density $\Sigma$, velocity dispersion $V_m$
(including the orbital motions of the stars and the turbulent
and thermal motions of the interstellar particles), RMS escape 
velocity $V_e$, and crossing time $\tau_c$.
For simplicity, we neglect rotation, magnetic support, and 
external pressure (but see Section 3). 
Then the properties of a protocluster are related by 
$V_m^2 = 0.4 GM/R_h$, $V_e = 2V_m$, $\tau_c = R_h/V_m$
\citep{spitzer87a} and $\Sigma \approx (M/2)/(\pi R_h^2)$.
We also assume that the sizes and masses of protoclusters 
are correlated, with a power-law trend, $R_h \propto M^{\alpha}$.

\begin{figure}
\plotone{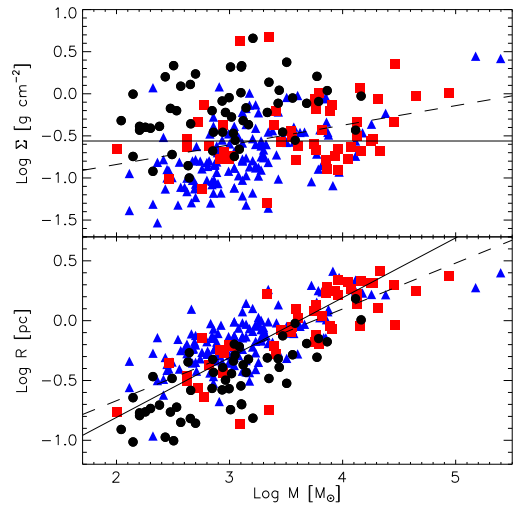}
\caption{
\label{msplot}
Surface density $\Sigma$ and radius $R$ plotted against mass $M$ for 
star-forming molecular clumps from measurements by Shirley et al. 
(2003; circles, CS emission), Fa{\'u}ndez et al. (2004; triangles,
dust emission), and Fontani et al. (2005; squares, C$^{17}$O and dust
emission).
We exclude clouds with $M < 100 M_\odot$, since they cannot form clusters.
The lines are least-squares regressions ($\log R$ against $\log M$) with 
$\alpha = 0.5$ fixed (solid) and $\alpha = 0.38 \pm 0.023$ (dashed).
The true uncertainty on $\alpha$ is undoubtedly larger than the quoted 
one-sigma error.} 
\end{figure}

In Figure \ref{msplot}, we plot $\Sigma$ and $R_h$ against $M$ 
for star-forming molecular clumps in the Milky Way, based on 
measurements of CS, C$^{17}$O, and 1.2 mm dust emission in three 
independent surveys \citep{shirley03, faundez04, fontani05}.
These clumps were selected for their star-formation activity
(water masers, IRAS colors), not their surface density.
Evidently, there is a strong correlation between $R_h$ and $M$,
and almost none between $\Sigma$ and $M$, corresponding to 
$\alpha \approx 1/2$.
The typical surface density is close to the value $\Sigma 
\sim 1$~g~cm$^{-2}$ expected from theory \citep{mckee03a, 
krumholz07a, krumholz08a}.\footnote{For reference, the
\citet{larson81} relation for CO-selected clouds corresponds
to a much lower surface density, $\Sigma \sim 0.02$~g~cm$^{-2}$.}
We assume that the Milky Way relations also hold in other
galaxies and extend up to $\sim 10^6 M_\odot$, although it is 
conceivable that they break down above $\sim 10^5 M_\odot$.
Indeed, \citet{baumgardt08a} and \citet{parmentier08a} assume that 
$R_h$ is not correlated with $M$ (corresponding to $\alpha = 0$), 
based on observations of gas-free clusters \citep[e.g.][]{murray09b}. 
However, since ISM removal necessarily occurs during the earlier,
gas-dominated phase, $\alpha \approx 1/2$ seems more appropriate  
in the present context.
As we show here, $\alpha \approx 1/2$ is also needed to reconcile
the observed mass functions of molecular clouds and star clusters.

The rates of energy and momentum input are proportional to the 
stellar mass\footnote{This is a good approximation for all 
feedback mechanisms except protostellar outflows, which inject 
energy and momentum in proportion to the star formation rate. Outflows, 
however, are non-dominant in massive protoclusters; see Table 1.}: 
$\dot{E} \propto {\cal E} M$ and $\dot{P} \propto {\cal E} M$. 
We assume that the timescale for ISM removal is a few crossing times:
$\Delta t \sim (1-10) \times \tau_c$ \citep{elmegreen00, 
elmegreen07, hartmann01, tan06a, krumholz07e}.
Thus, the total energy and momentum input are 
$E \approx \dot{E}{\Delta}t \propto {\cal E} M R_h/V_m$ and
$P \approx \dot{P}{\Delta}t \propto {\cal E} M R_h/V_m$.
These reach the critical values needed to remove the ISM, 
$E_{\rm crit} = {\onehalf} M V_e^2$ and $P_{\rm crit} = M V_e$, 
for 
\begin{subequations}
\begin{eqnarray}
{\cal E} & \propto & V_e^3/R_h \propto M^{(3 - 5\alpha)/2}
\,\,\,\,\,\,\,\,\,\,
{\rm (energy-driven)},
\label{effic:energy} 
\\
{\cal E} & \propto & V_e^2/R_h \propto M^{1 - 2\alpha}
\,\,\,\,\,\,
{\rm (momentum-driven)}. 
\label{effic:momentum}
\end{eqnarray}
\end{subequations}
For $\alpha=1/2$, the efficiency has little or no dependence on mass:
${\cal E} \propto M^{1/4}$ in the energy-driven regime, ${\cal E}
= {\rm constant}$ in the momentum-driven regime. 
For $\alpha=0$, the variation is much stronger: ${\cal E} \propto M^{3/2}$ 
and ${\cal E}\propto M$, respectively.
These relations are valid for ${\cal E} \la 0.5$.

Any dependence of ${\cal E}$ on $M$ will cause the mass functions of 
star clusters $\psi_*(M_*)$ and molecular clouds $\psi(M)$ to have 
different shapes. 
For the moment, we confine our attention to clusters young enough 
to be easily recognizable even if they are unbound and dispersing.
This limit is $\sim 10^7$~yr for extragalactic clusters such as 
those in the Antennae \citep{fall05a}.
In this case, the mass functions of the clusters and clouds are
related by $\psi_*(M_*)dM_* \propto \psi(M)dM$ (with a coefficient 
greater than unity if several clusters form within each cloud).
For $\psi(M) \propto M^{\beta}$ and ${\cal E} \propto M^{\gamma}$,
we have $\psi_*(M_*) \propto M_*^{\beta_*}$ with 
$\beta_* = (\beta - \gamma)/(1 + \gamma)$. 
Equations (\ref{effic:energy}) and (\ref{effic:momentum}) then imply
\begin{subequations}
\begin{eqnarray}
\beta_* & = & \frac{2\beta + 5\alpha - 3}{5(1 - \alpha)} 
\,\,\,\,\,\,\,\,\,\,\,\,\,\,\,\,\,\,\,\,\,\,\,
{\rm (energy-driven)}, 
\label{beta1e}
\\
\beta_* & = & \frac{\beta + 2\alpha - 1}{2(1 - \alpha)}
\,\,\,\,\,\,\,\,\,\,\,\,\,\,
{\rm (momentum-driven)}. 
\label{beta1p}
\end{eqnarray}
\end{subequations}
These expressions give $\beta_* = \beta$ for $\alpha = 3/5$ and 1/2, 
respectively.
Thus, the similarity of the mass functions of clusters 
and clouds ($\beta_* \approx \beta$) {\it requires} 
that the latter have approximately constant mean surface 
density ($0.5 \la \alpha \la 0.6$), no matter what type of
feedback is involved.

Before proceeding, we make a small correction.
For clouds, the observed mass function $\psi_o(M)$ represents the 
true mass function at formation $\psi(M)$ (i.e., the birthrate) 
weighted by the lifetime: $\psi_o(M) \propto \psi(M)\tau_l(M)$.
We assume, as before, that lifetime is proportional to crossing 
time: $\tau_l \propto \tau_c \propto M^{(3\alpha - 1)/2}$.
Then the exponents of the true and observed mass functions are
related by $\beta = \beta_o - (3\alpha - 1)/2$.
Inserting this into Equations (\ref{beta1e}) and (\ref{beta1p}), 
we obtain 
\begin{subequations}
\begin{eqnarray}
\beta_* & = & \frac{2(\beta_o + \alpha - 1)}{5(1 - \alpha)}
\,\,\,\,\,\,\,\,\,\,\,\,\,\,\,\,\,\,\,\,\,\,\,\,\,\,
{\rm (energy-driven)}, 
\label{betastar:energy}
\\
\beta_* & = & \frac{2\beta_o + \alpha - 1}{4(1 - \alpha)}
\,\,\,\,\,\,\,\,\,\,\,\,\,\,\,\,\,\,
{\rm (momentum-driven)}. 
\label{betastar:momentum}
\end{eqnarray}
\end{subequations}

We now evaluate Equations (\ref{betastar:energy}) and 
(\ref{betastar:momentum}) with $\beta_o = -1.7$, the
observed exponent of the mass function of molecular clouds 
\citep{rosolowsky05b, reid06b, munoz07a, wong08a, fukui08a}.
For constant mean surface density ($\alpha = 1/2$), we find
$\beta_* = -1.8$ in the energy-driven regime and $\beta_* = -2.0$ 
in the momentum-driven regime.
These predictions agree nicely with the observed exponents of
the mass functions of young star clusters, $\beta_* \approx -2.0$
(with typical uncertainty $\Delta\beta_* \approx 0.2$). 
Our model is clearly idealized, but the scalings, and thus the agreement 
between the predicted and observed $\beta_*$, should be robust.
For constant size ($\alpha = 0$), however, we find $\beta_* = -1.1$ 
in both the energy-driven and momentum-driven regimes, in definite
conflict with observations.

The mass function of star clusters older than $\sim 10^7$~yr 
depends on the proportion that remain gravitationally bound.
This in turn depends on the efficiency of star formation ${\cal E}$
and the timescale for ISM removal $\Delta t$ relative to the crossing 
time $\tau_c$.
Both analytical arguments and $N$-body simulations indicate that
young clusters lose most of their stars for ${\cal E} \la 0.3$ and
${\Delta t} \ll \tau_c$ but retain most of them for ${\cal E} \ga 0.5$
or ${\Delta t} \gg \tau_c$ \citep{hills80, kroupa01b, kroupa02a, 
baumgardt07a}.
Thus, as long as ${\cal E}$ and ${\Delta t}/\tau_c$ are, on average,
independent of $M$, as they are for protoclusters with constant 
mean surface density ($\alpha = 1/2$) and momentum-driven feedback, 
ISM removal will not alter the shape of the mass function 
(although its amplitude will decline). 
This is consistent with the observed exponents $\beta_* \approx -2.0$
for clusters both younger and older than $10^7$~yr in the Antennae
and LMC \citep{zhang99b, fall09a, chandar09a}.

In all other cases, ${\cal E}$ increases with $M$, and a higher 
proportion of low-mass clusters is disrupted, causing a flattening 
or a bend at ${\cal E} \approx 0.3 - 0.5$ in the mass function.
The exact shape depends on $\Delta t/\tau_c$, clumpiness within
protoclusters, and other uncertain factors.
If the efficiency has a weak dependence on mass, as it does for 
constant mean surface density ($\alpha = 1/2$) and energy-driven
feedback (${\cal E} \propto M^{1/4}$), the predicted $\beta_*$
might be marginally consistent with observations over a limited 
range of masses ($10^4 M_{\odot} \la M \la 10^6 M_{\odot}$).
However, for constant size ($\alpha = 0$), the variations are
so strong (${\cal E} \propto M^{3/2}$ and ${\cal E} \propto M$) 
that we expect major differences between the mass functions of 
clusters younger and older than $10^7$~yr, in clear contradiction 
with observations.

Our simple analytical model agrees, at least qualitatively, with 
the numerical calculations by \citet{baumgardt08a} and 
\citet{parmentier08a}. 
They present results for energy-driven feedback by supernovae in
protoclusters with uncorrelated sizes and masses. 
In some cases, they find a bend in the mass function of young clusters 
at $M \sim 10^5 M_{\odot}$, while in others, they find a flattened power 
law with $\beta_* \approx -1$ (see Figure 4 of \citealt{baumgardt08a}).
As we have already noted, these results are expected for $\alpha = 0$, 
and they are inconsistent with the observed mass functions of young 
clusters.

\section{Star Formation Efficiency}
\label{sec:efficiency}

\begin{deluxetable*}{ccccc}
\tabletypesize{\scriptsize}
\tablecaption{Feedback Mechanisms}
\tablehead{ \colhead{Mechanism} & \colhead{Type} & \colhead{Limitation} & 
\colhead{Threshold\tablenotemark{\dag}} & 
\colhead{Evaluated\tablenotemark{\dag}}  }
\startdata 
Supernovae & Energy & Too late &  $\tau_c \approx 1.8$ Myr & 
$\Sigma_0 \approx 0.022 M_4^{1/3} $ \\
Main-sequence winds &  Either\tablenotemark{a} & Relatively weak\tablenotemark{a} 
& Never & \nodata \\ 
Protostellar outflows & Momentum & Confined in massive clusters\tablenotemark{b} 
& $V_e \approx 7$ km s$^{-1}$ & $\Sigma_0 \approx 0.17 M_4^{-1}$ \\ 
Photoionized gas & Momentum & Crushed by $P_{\rm rad}$\tablenotemark{c} & 
$S_{49} \approx 21 R_h/\mbox{pc}$ & $\Sigma_0 \approx 0.15 M_4^{-1}$ \\
Radiation pressure & Momentum &\nodata &  Equations (\ref{efficiency}) and
(\ref{Sigma_crit}) & $\Sigma_0 \approx 1.2$
\enddata
\tablenotetext{\dag}{Parameters required for ${\cal E} = 0.5$. 
Evaluations assume a fully-sampled stellar IMF. Notation:
$S_{49} \equiv S/10^{49}$~s$^{-1}$ (ionization rate), 
$M_4 \equiv M/10^4 M_\odot$, $\Sigma_0 \equiv \Sigma/$g~cm$^{-2}$.}
\tablenotetext{a}{Stellar winds are energy-driven and dominant if trapped, 
but are expected to leak, making them momentum-driven and weak.} 
\tablenotetext{b}{Based on Equation (55) of \citet{matzner00}, 
updated with $f_w v_w = 80$ km s$^{-1}$ \citep{matzner07}.}
\tablenotetext{c}{Based on Equation (4) of \citet{krumholz09d} 
for the blister case, with the coefficient reduced by a factor 
of $2.2^2$ to correct an error in the published paper and
updated with $\langle L/M_*\rangle = 1140 \lsun \msun^{-1}$ and 
$\langle S/M_*\rangle = 6.3\times 10^{46}$ s$^{-1}$ $\msun^{-1}$ 
\citep{murray09c}.}
\label{Table}
\end{deluxetable*}

We now consider five specific feedback mechanisms: 
supernovae, main-sequence winds, protostellar outflows, 
photoionized gas, and radiation pressure. 
For the first four, we review results from the literature. 
Supernova feedback begins only after the $>3.6$ Myr lifetimes of 
massive stars. 
Unless turbulence within a protocluster is maintained by feedback
or external forcing, stars would form rapidly and consume its ISM,
with ${\cal E} \rightarrow 1$ in $1-2$ crossing times.
This implies that supernovae can dominate only for $2\tau_c \ga 
3.6$ Myr unless another mechanism somehow keeps ${\cal E}$ small 
without expelling much ISM \citep{krumholz09d}.
However, even in this contrived situation, supernovae would play
only a secondary role.
Main-sequence winds are not effective if their energy is able to 
leak out of the bubbles they blow \citep{harper-clark09a}. 
As a result of this leakage, winds simply provide an order-unity 
enhancement to radiation pressure \citep{krumholz09d}.
Protostellar outflows can only remove the ISM from protoclusters
with escape velocities below about 7 kms$^{-1}$ \citep{matzner00}.
Photoionized gas is important as a feedback mechanism only when
its pressure exceeds that of radiation throughout most of an 
\hii\ region.
This in turn requires that the \hii\ region be larger than the
radius $r_{\rm ch}$ at which $P_{\rm rad} = P_{\rm gas}$, a condition
harder to satisfy in massive, compact protoclusters
\citep{krumholz09d}.

\begin{figure}
\plotone{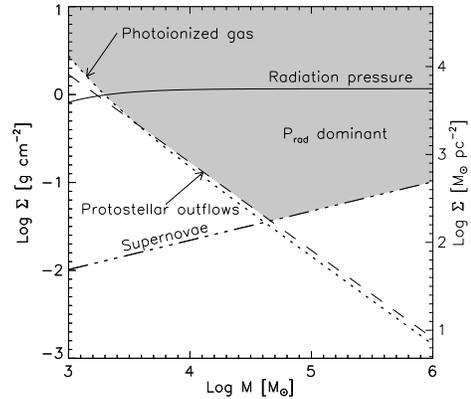}
\caption{
\label{Fig:ejection}
Feedback in protoclusters of mean surface density $\Sigma$ and mass $M$.
Radiation pressure is the dominant mechanism throughout the shaded region.
The lines show where each mechanism alone achieves ${\cal E} = 0.5$. 
These allow for partial sampling of the stellar IMF and hence differ 
slightly from the power laws in Table \ref{Table} (noticeable only for
$M \la 10^4 M_\odot$).}
\end{figure}

We summarize these results in Table \ref{Table} and Figure \ref{Fig:ejection}. 
As the plot shows, the mechanisms discussed thus far are relatively ineffective
in protoclusters with $M \ga 10^4$ $\msun$ and $\Sigma \ga 0.1$ g cm$^{-2}$. 
We therefore turn to radiation pressure.
This would be an energy-driven feedback mechanism if all photons, even those 
re-radiated by dust grains, remained trapped within a protocluster.
However, this is possible only if the protocluster is so dense and 
smooth that the covering fraction seen from its center exceeds $\sim90$
\% in the infrared \citep{krumholz09d}.
More realistically, the protocluster would be porous enough that photons 
could escape after only a few interactions with dust grains, and radiation 
pressure would then be a momentum-driven feedback mechanism. 
The following analysis extends that of 
\citet{elmegreen83}, \citet{scoville01}, \citet{thompson05},
\citet{krumholz09d}, and \citet{murray09a}.

We consider an idealized, spherical cloud of mass $M$ and outer 
radius $R$, with an internal density profile $\rho\propto r^{-k}$
(hence $R_h=2^{-1/(3-k)} R$). 
Radiation from young stars near the center of the cloud ionizes the gas 
and drives the expanding outer shell of this \hii\ region. 
After a time $t$, the momentum imparted to the shell is 
$ p_s = f_{\rm trap} L t/c$, where $L$ is the stellar luminosity 
(assumed constant for simplicity), and $f_{\rm trap}\sim 2-5$ 
accounts for assistance from main-sequence winds and incomplete leakage 
of starlight and wind energy \citep{krumholz09d}.  
Neglecting gravity for the moment, the velocity and radius of the
shell are related by $v_s = \eta r_s/t$ with $\eta= 2/(4-k)$.  
Thus, when the shell reaches the cloud surface ($r_s = R$), it has 
swept up all the remaining ISM, with mass $M_g=(1-{\cal E})M$, and 
has a velocity given by
\begin{equation}
v_s^2(R) =\frac{\eta  f_{\rm trap} L R}{c (1 - {\cal E}) M}. 
\end{equation} 

We specify the condition for ISM removal against gravity of the
protocluster by $v_s^2(R) = \alpha_{\rm crit} G M/(5 R)$, where 
$\alpha_{\rm crit}$ is a parameter of order unity that accounts 
for magnetic support and other uncertain factors (discussed below).
The required luminosity, from equation (4), is
\begin{equation} \label{L_crit} 
L = \frac{\alpha_{\rm crit} G c (1-{\cal E})M^2}{5 \eta f_{\rm trap} R^2}.   
\end{equation} 
The fundamental scaling $L\propto (M/R)^2 \propto V_m^4$ arises here in 
the same way it does for the growth of supermassive black holes and 
galactic spheroids \citep{fabian99, king03, murray05a}. 
Rewriting Equation (\ref{L_crit}) in terms of $\Sigma = M/(\pi R^2)$
and $M_* = {\cal E} M$ and solving for ${\cal E}$, we obtain our basic
result 
\begin{equation} \label{efficiency} 
{\cal E} =  \frac{\Sigma }{\Sigma +  \Sigma_{\rm crit}},
\end{equation} 
with 
\begin{equation}\label{Sigma_crit}
\Sigma_{\rm crit}  =\frac{ 5 \eta f_{\rm trap}  (L/M_*) }{ \pi \alpha_{\rm crit} G c} 
\approx 1.2 \left(\frac{f_{\rm trap}}{\alpha_{\rm crit}}\right) {\rm g\,cm^{-2}}. 
\end{equation}
The coefficient in the last equation is based on $\eta = 2/3$ and 
$L/M_* = 1140 L_\odot/M_\odot$ (see notes to Table \ref{Table}). 
Regardless of the exact value of $f_{\rm trap}/\alpha_{\rm crit}$, we note that 
${\cal E}$ depends on $M$ and $R$ only through $\Sigma$.
Thus, when $\Sigma$ is constant, ${\cal E}$ is independent of $M$, and the mass 
functions of clusters and clouds have the same exponent ($\beta_* = \beta
\approx \beta_o$). 
 
\begin{figure}
\plotone{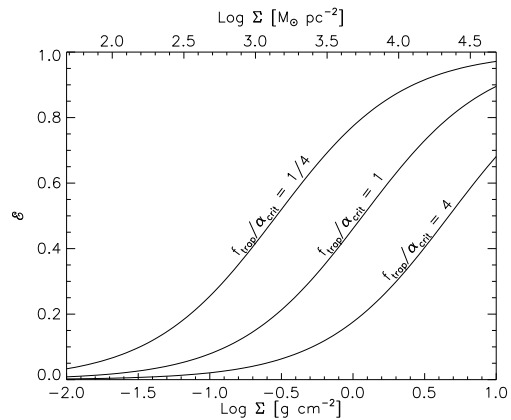}
\caption{
\label{sfeplot}
Star formation efficiency ${\cal E}$ as a function of mean surface density 
$\Sigma$, computed from Equations (\ref{efficiency}) and (\ref{Sigma_crit}) 
with the indicated values of $f_{\rm trap}/\alpha_{\rm crit}$.}
\end{figure}

Figure \ref{sfeplot} shows ${\cal E}(\Sigma)$ computed from Equations 
(\ref{efficiency}) and (\ref{Sigma_crit}). 
Clearly, ${\cal E}$ increases monotonically with $\Sigma$ from 0 to 1, reaching 
${\cal E} = 0.3$ for $\Sigma \sim 0.5 (f_{\rm trap}/\alpha_{\rm crit})$ g cm$^{-2}$.
We expect $f_{\rm trap} \sim \alpha_{\rm crit} \sim$~2--5.
The escape velocity from the surface of an unmagnetized cloud corresponds to 
$\alpha_{\rm crit}=10$, while the internal velocity dispersion, possibly sufficient
for some ISM removal, corresponds to $\alpha_{\rm crit} \approx 1.3$.
A shell driven by a constant force requires $\alpha_{\rm crit}=2.3$ (for $k=1$; 
see Equation (A17) of \citet{matzner00}). 
We consider $\alpha_{\rm crit}\approx 2$ to be plausible; certainly a 
protocluster boils violently and loses mass rapidly using this condition.  
Our intent here is not to make a detailed comparison between the model 
and observations. Given the simplicity of the former and the uncertainties 
in the latter, it is gratifying that they agree even roughly with each other.

\section{Conclusions}
\label{sec:conclusion}
 
This Letter contains two main results.
The first is the relation between the power-law exponents of the
mass functions of molecular clouds and young star clusters, 
$\beta_o$ and $\beta_*$, in the limiting regimes in which 
stellar feedback is energy-driven and momentum-driven, Equations
(\ref{betastar:energy}) and (\ref{betastar:momentum}), which 
bracket all realistic cases. 
The predicted $\beta_*$ depends significantly
on the initial size-mass relation of the protoclusters.
We find good agreement between the predicted and observed $\beta_*$,
especially for momentum-driven feedback, for $\Sigma \propto M/R_h^2 
\approx {\rm constant}$, 
the relation indicated by observations of gas-dominated protoclusters.
In this case, the star formation efficiency is independent of 
protocluster mass, ensuring that the fraction of clusters that 
remain gravitationally bound following ISM removal
is also independent of mass.

The second main result is an estimate of the star formation efficiency 
in protoclusters regulated by radiation pressure, 
Equations (\ref{efficiency}) and (\ref{Sigma_crit}).
This is likely to be the dominant feedback process in massive 
protoclusters.
We show that ${\cal E}$ depends on $M$ and $R_h$ only through the
mean surface density $\Sigma$, which in turn guarantees 
consistency between the observed power-law exponents of the mass 
functions of molecular clouds and young star clusters according to 
our general relations.
For $\Sigma \sim 1$~g~cm$^{-2}$, we estimate ${\cal E} \sim 0.3$,
in satisfactory agreement with observations.

\acknowledgements
We thank Bruce Elmegreen, Chris McKee, Dean McLaughlin, Norm Murray, 
John Scalo, Nathan Smith, and the referee for helpful comments.
We are grateful for research grants from NASA (SMF, MRK, CDM), NSF 
(MRK), Sloan Foundation (MRK), NSERC (CDM), and Ontario MRI (CDM).


\end{document}